\documentclass[10pt,twoside]{Lowx2011}
\usepackage{epsf,amsmath}
\usepackage{graphicx}

\newcommand{\beq}{\begin{equation}}
\newcommand{\eeq}{\end{equation}}
\newcommand{\bea}{\begin{eqnarray}}
\newcommand{\eea}{\end{eqnarray}}

\begin{document}

\title{HARD INCLUSIVE PRODUCTION OF A PAIR OF RAPIDITY-SEPARATED HADRONS
IN PROTON COLLISIONS}

\author{Dmitry Yu. Ivanov$^1$ and \underline{Alessandro Papa}$^2$\\ \\
${}^1$ {\sl Sobolev Institute of Mathematics and Novosibirsk State
University,} \\
{\sl 630090 Novosibirsk, Russia}\\ \\
${}^2$ {\sl Dipartimento di Fisica, Universit\`a della Calabria,} \\
{\sl and Istituto Nazionale di Fisica Nucleare, Gruppo collegato di Cosenza,}\\
{\sl I-87036 Arcavacata di Rende, Cosenza, Italy}\\ \\
E-mail: d-ivanov@math.nsc.ru, papa@cs.infn.it}

\maketitle

\begin{abstract}
\noindent 
We discuss the process $p+p\to h_1+h_2+X$, where the identified hadrons 
$h_1$ and $h_2$ have large transverse momenta and are produced in high-energy 
proton-proton collisions with a large rapidity gap between them.
In this case the (calculable) hard part of the reaction receives large 
higher order corrections $\sim \alpha^n_s\ln^n \Delta y$, which can be
accounted for in the BFKL approach. Specifically, we describe in the 
next-to-leading order the calculation of the vertex (impact-factor) for the 
inclusive production of the identified hadron.
\end{abstract}



\markboth{\large \sl \hspace*{0.25cm}D.Yu. Ivanov \& \underline{A. Papa}
\hspace*{0.25cm} Low-$x$ Meeting 2011} {\large \sl \hspace*{0.25cm} 
HARD INCLUSIVE PRODUCTION OF A PAIR ...}

\section{Introduction}

The process under consideration is
\[
{\rm proton}(p_1)   \, + \, {\rm proton}(p_2) \, \to \, 
{\rm hadron}_1(k_1) \, + \, {\rm hadron}_2(k_2) \, + \, X \;.
\]
Introducing the Sudakov decomposition for the momentum of each 
identified hadron,
\[
k_h= \alpha_h p_1+ \frac{\vec k_h^2}{\alpha_h s}p_2+k_{h\perp} \;, 
\;\;\;\;\; k_{h\perp}^2=-\vec k_h^2 \;,\;\;\;\;\; s=2 p_1\cdot p_2 \;,
\]
we assume that hadrons' transverse momenta are large, 
$\vec k_1^{\:2}\sim \vec k_2^{\:2} \gg \Lambda_{\rm QCD}^2$, so that
perturbative QCD is applicable. Moreover, we consider the high-energy limit
$s=2 p_1\cdot p_2 \gg \vec k_{1,2}^{\:2}$, which opens the way to the
BFKL~\cite{BFKL} resummation.

Let us briefly remind the basics of the BFKL approach. In the Regge limit 
($s \to \infty$, $t$ not growing with $s$), the total cross section 
$A + B \to X$ can be written as (see, for instance,~\cite{FF98})
\[
\sigma_{AB}=\frac{1}{(2\pi)^{D-2}}\!\int\!\frac{d^{D-2}\vec q_1}{\vec
q_1^{\,\, 2}}\Phi_A(\vec q_1,s_0)\!\int\!
\frac{d^{D-2}\vec q_2}{\vec q_2^{\,\,2}} \Phi_B(-\vec q_2,s_0)
\!\int\limits^{\delta +i\infty}_{\delta
-i\infty}\!\frac{d\omega}{2\pi i}\left(\frac{s}{s_0}\right)^\omega
G_\omega (\vec q_1, \vec q_2)\,.
\]
This factorization is valid both in the leading logarithmic approximation 
(LLA), which means resummation of all terms $(\alpha_s\ln s)^n$, and in the 
next-to-LLA (NLA), which means resummation of all terms 
$\alpha_s(\alpha_s\ln s)^n$. The Green's function $G_\omega$ is 
process-independent and is determined through the BFKL equation
in $D=4+2\epsilon$ dimensions,
\[
\omega \, G_\omega (\vec q_1,\vec q_2)  =\delta^{D-2} (\vec q_1-\vec q_2)
+\int d^{D-2}\vec q \, K(\vec q_1,\vec q) \,G_\omega (\vec q, \vec q_1) \;,
\]
whose kernel is known in the NLA both for forward 
scattering (i.e. for $t=0$ and color singlet in the 
$t$-channel)~\cite{FL+CC98} and for any fixed (not growing with 
energy) momentum transfer $t$ and any possible two-gluon color state in the 
$t$-channel~\cite{FF05}. As for the process-dependent impact factors (IFs) 
$\Phi_{A,B}$, only very few have been calculated in the NLA. 

\begin{figure}[tb]
\centering
\includegraphics[scale=0.6]{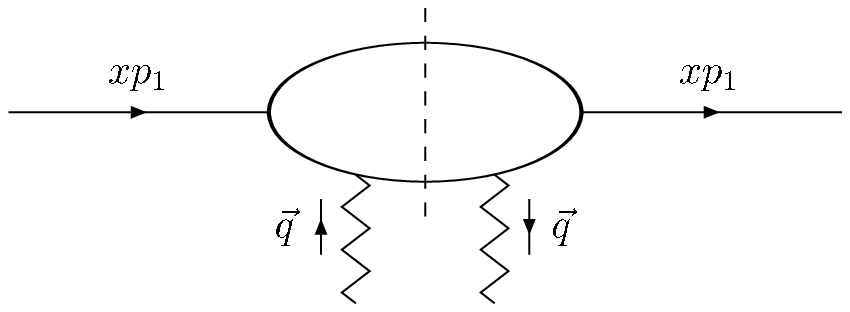}
\hspace{2cm}
\includegraphics[scale=0.6]{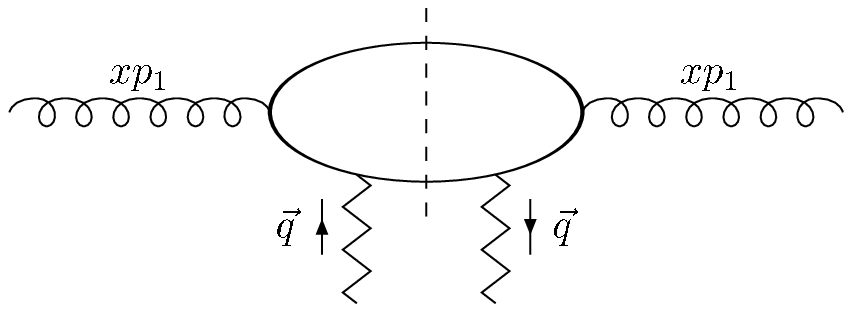}
\vspace{-0.2cm}
\caption{Diagrammatic representation of the forward quark (left) and gluon 
(right) impact factor.}
\label{fig:if}
\end{figure}

The starting point for the calculation in the NLA of the IF
relevant for the process under consideration is provided by
the IFs for colliding partons~\cite{FFKP99+Cia} (see Fig.~\ref{fig:if}).
We observe that for the LLA IF, there can be only a 
one-particle intermediate state, whereas for the NLA IF,
we can have virtual corrections to the one-particle intermediate state,
but also real particle production, with a two-particle intermediate state.

\begin{figure}[bt]
\centering
\includegraphics[scale=0.6]{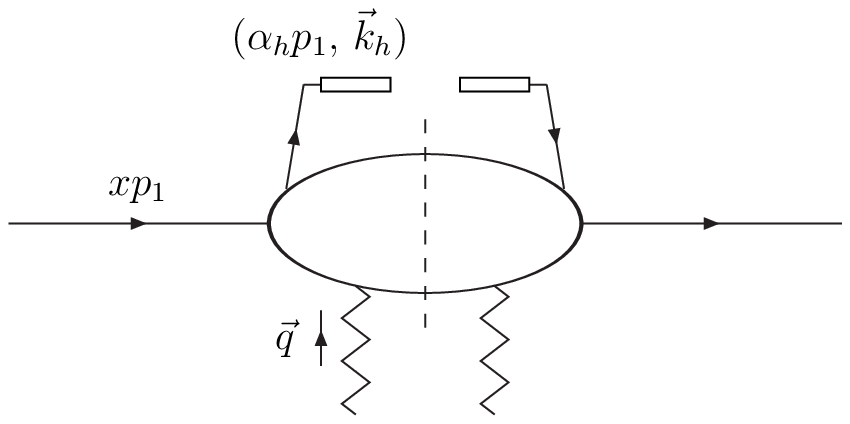}
\hspace{1.8cm}
\includegraphics[scale=0.6]{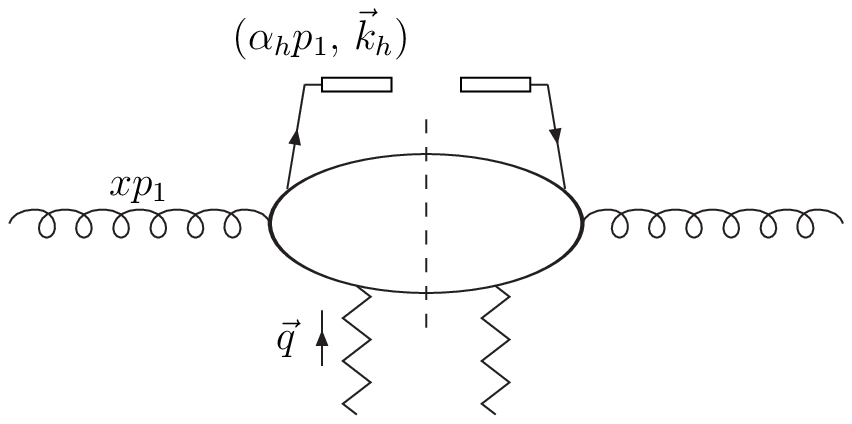}
\vspace{-0.2cm}
\caption{Diagrammatic representation of the vertex for the identified
hadron production for the case of incoming quark (left) or gluon (right).}
\label{fig:vertex}
\end{figure}

Here are the steps of the calculation:

i) ``open'' one of the integrations over the phase space of the 
intermediate state to allow one parton to fragment into a given hadron
(see Fig.~\ref{fig:vertex}); 

ii) use QCD collinear factorization,
\[
\sum_{a=q,\bar q} f_a \otimes ({\rm quark \ vertex}) \otimes D_a^h
\; + \; f_g \otimes ({\rm gluon \ vertex}) \otimes D_g^h\;,
\]

iii) project onto the eigenfunctions of the LLA BFKL kernel 
($(\nu,n)$-representation),
\[
\Phi(\nu,n)=\int d^2\vec q \,\frac{\Phi(\vec q)}{\vec q^{\,\, 2}}\frac{1}{\pi
\sqrt{2}}\left(\vec q^{\,\, 2}\right)^{\gamma-\frac{n}{2}} 
\left(\vec q \cdot \vec l \,\, \right)^n \;,
\;\;\;\;\;
\gamma=i\nu-\frac{1}{2}\;,\;\;\;\;\;\vec l^{\:2}=0 \;,
\]
which is convenient for the numerical convolution with BFKL Green's function.

\begin{figure}[tb]
\centering
\includegraphics[scale=0.6]{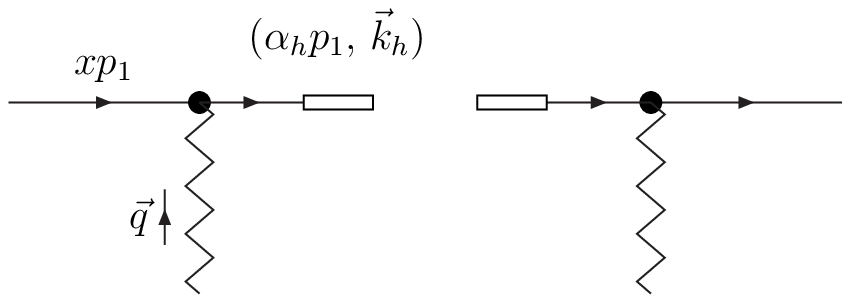}
\hspace{2cm}
\includegraphics[scale=0.6]{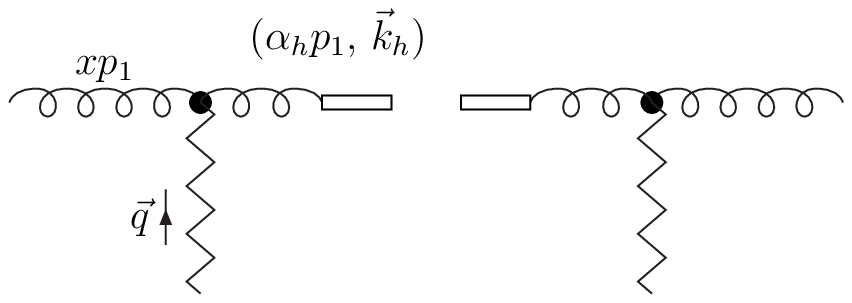}
\vspace{-0.2cm}
\caption{Diagrammatic representation of the LLA vertex for the case of 
incoming quark (left) and gluon (right).}
\label{fig:vertex_LLA}
\end{figure}

\section{The impact factor in the LLA}

The starting point is given by the ``inclusive'' LLA parton IFs:
\[
\Phi_q=g^2\frac{\sqrt{N^2-1}}{2N}\;,\;\;\;\;\;\Phi_g=\frac{C_A}{C_F}\Phi_q\;,
\;\;\;\;\;\;\;\;\;\;C_A=N\;, \;\;\; C_F=\frac{N^2-1}{2N}\;.
\]
Here the step i) means simply to introduce a delta function (see 
Fig.~\ref{fig:vertex_LLA}). Then, QCD collinear factorization leads to
\[
\frac{d\Phi^h}{\vec q^{\,\, 2}}=\Phi_q \,d\alpha_h \frac{d^{2+2\epsilon}
\vec k}{\vec k^{\, 2}}\int\limits^1_{\alpha_h} \frac{dx}{x}
\, \delta^{(2+2\epsilon)}(\vec k-\vec q)
\left(\frac{C_A}{C_F}f_g(x)
D^h_g\left(\frac{\alpha_h}{x}\right)+\sum_{a=q,\bar q} f_a(x)
D^h_a\left(\frac{\alpha_h}{x}\right)\right)\;.
\]

\section{The impact factor in the NLA}

Collinear singularities in NLA are to be removed by PDFs' and FFs'
renormalization:
\begin{eqnarray*}
&f_q(x)=f_q(x,\mu_F)-\frac{\alpha_s}{2\pi}\left(\frac{1}{\hat \epsilon}
+\ln\frac{\mu_F^2}{\mu^2}\right)
\int\limits^1_x\frac{dz}{z}\left[P_{qq}(z)f_q(\frac{x}{z},\mu_F)
+P_{qg}(z)f_g(\frac{x}{z},\mu_F)\right]& \\
&f_g(x)=f_g(x,\mu_F)-\frac{\alpha_s}{2\pi}\left(\frac{1}{\hat \epsilon}
+\ln\frac{\mu_F^2}{\mu^2}\right)
\int\limits^1_x\frac{dz}{z}\left[P_{gq}(z)f_q(\frac{x}{z},\mu_F)
+P_{gg}(z)f_g(\frac{x}{z},\mu_F)\right]\,,& \\
&D^h_q(x)=D^h_q(x,\mu_F)-\frac{\alpha_s}{2\pi}\left(\frac{1}{\hat \epsilon}
+\ln\frac{\mu_F^2}{\mu^2}\right)
\int\limits^1_x\frac{dz}{z}\left[D^h_q(\frac{x}{z},\mu_F)P_{qq}(z)
+D^h_g(\frac{x}{z},\mu_F)P_{gq}(z)\right]& \\
&D^h_g(x)=D^h_g(x,\mu_F)-\frac{\alpha_s}{2\pi}\left(\frac{1}{\hat \epsilon}
+\ln\frac{\mu_F^2}{\mu^2}\right)
\int\limits^1_x\frac{dz}{z}\left[D^h_q(\frac{x}{z},\mu_F)P_{qg}(z)
+D^h_g(\frac{x}{z},\mu_F)P_{gg}(z))\right]\,,&
\end{eqnarray*}
where $P_{ij}$ are the Altarelli-Parisi splitting functions and
$\frac{1}{\hat \epsilon}=\frac{1}{\epsilon}+\gamma_E-\ln (4\pi)\approx 
\frac{\Gamma(1-\epsilon)}{\epsilon (4\pi)^\epsilon}$.
This leads to the following collinear counterterms:
\[
\frac{\pi\sqrt{2}\, \vec k^{\, 2}}{\Phi_q}
\frac{d\Phi^h(\nu,n)|_{\rm{coll.\ c.t.}}}
{d\alpha_h d^{2+2\epsilon}\vec k}
= -  \frac{\alpha_s}{2\pi}\left(\frac{1}{\hat \epsilon}+\ln\frac{\mu_F^2}
{\mu^2}\right)\int\limits^1_{\alpha_h} \frac{dx}{x}
\int\limits^1_{\frac{\alpha_h}{x}} \frac{dz}{z}
\left(\vec k^{\,\,2} \right)^{\gamma-{n \over 2}}
\left(\vec k \cdot \vec l \,\, \right)^n 
\]
\[
\times \left[  (1+z^{-2\gamma})P_{qq}(z)\sum_{a=q,\bar q}f_a(x)
D_a^h\left(\frac{\alpha_h}{xz}\right)+ \left(\frac{C_A}{C_F}
+z^{-2\gamma}\right) P_{gq}(z)\sum_{a=q,\bar q}f_a(x)
D_g^h\left(\frac{\alpha_h}{xz}\right)\right. 
\]
\[
\left.
+(1+z^{-2\gamma})\frac{C_A}{C_F}P_{gg}(z) f_g(x)
D_g^h\left(\frac{\alpha_h}{x z}\right)+ \frac{C_A}{C_F}
\left(\frac{C_F}{C_A}+z^{-2\gamma}\right)P_{qg}(z) f_g(x)
\!\sum_{a=q,\bar q}D_a^h\left(\frac{\alpha_h}{x z}\right)\right].
\]

The other counterterm comes from the QCD coupling renormalization 
and reads
\[
\frac{\pi\sqrt{2}\, \vec k^{\, 2}}{\Phi_q}\frac{d\Phi(\nu,n)|_{\rm{charge
\ c.t.}}}
{d\alpha d^{2+2\epsilon}\vec k}= \frac{\alpha_s}{2\pi}
\left(\frac{1}{\hat \epsilon}+\ln\frac{\mu_R^2}{\mu^2}\right)
\left(\frac{11 C_A}{6}-\frac{n_f}{3}\right)
\]
\[
\times \int\limits^1_{\alpha_h}\frac{dx}{x}
\left( \frac{C_A}{C_F}f_g(x) D_g^h\left(\frac{\alpha_h}{x}\right)
+\sum_{a=q,\bar q}f_a(x) D_a^h\left(\frac{\alpha_h}{x}\right)\right)
\left(\vec k^{\,\,2} \right)^{\gamma-{n \over 2}}
\left(\vec k \cdot \vec l \,\, \right)^n\;.
\]

In the following we will use the following abbreviation
$\frac{\pi\sqrt{2}\, \vec k^{\, 2}}{\Phi_q}\frac{d\Phi^h(\nu,n)}
{d\alpha_h d^{2+2\epsilon}\vec k}\equiv I$.

\subsection{Quark-initiated subprocess}

We have first of all the virtual corrections to the one-particle 
intermediate state:
\[
I_q^V=-\frac{\alpha_s}{2\pi}\frac{\Gamma[1-\epsilon]}
{(4\pi)^\epsilon}\frac{1}{\epsilon}
\frac{\Gamma^2(1+\epsilon)}{\Gamma(1+2\epsilon)}  \int\limits^1_{\alpha_h}
\frac{dx}{x}\sum_{a=q,\bar q}f_a(x) D_a^h\left(\frac{\alpha_h}{x}
\right)\left(\vec k^{\,\,2}\right)^{\gamma+\epsilon-{n \over 2}}
\left(\vec k \cdot \vec l \,\, \right)^n
\]
\[
\times \left\{C_F\left(\frac{2}{\epsilon}-3\right)
-\frac{n_f}{3}+C_A\left(\ln\frac{s_0}{\vec k^2}+\frac{11}{6}\right)
\right\} + {\rm finite \ terms}\;.
\]

Then, we have to consider the ``real'' corrections from the 
quark-gluon intermediate state. The starting point is the ``inclusive''
quark IF,
\[
\Phi^{\{QG\}}=\Phi_q g^2\vec q^{\,\, 2}\frac{d^{2+2\epsilon} \vec k_1}
{(2\pi)^{3+2\epsilon}}\frac{d\beta_1}{\beta_1}\frac{[1+\beta_2^2
+\epsilon \beta_1^2]}{\vec k_1^{\, 2} \vec k_2^{\, 2}
(\vec k_2\beta_1-\vec k_1 \beta_2)^2} \left\{C_F \beta_1^2\vec k_2^{\, 2}
+C_A\beta_2\left(\vec k_1^{\,2}-\beta_1\vec k_1\cdot\vec q\right) \right\}\,,
\]
where $\beta_{1,2}$ and $\vec k_{1,2}$ are the relative longitudinal and 
transverse momenta of the gluon (quark) and $\beta_1+\beta_2=1$, 
$\vec k_1+\vec k_2=\vec q$.

\begin{figure}[tb]
\centering
\includegraphics[scale=0.6]{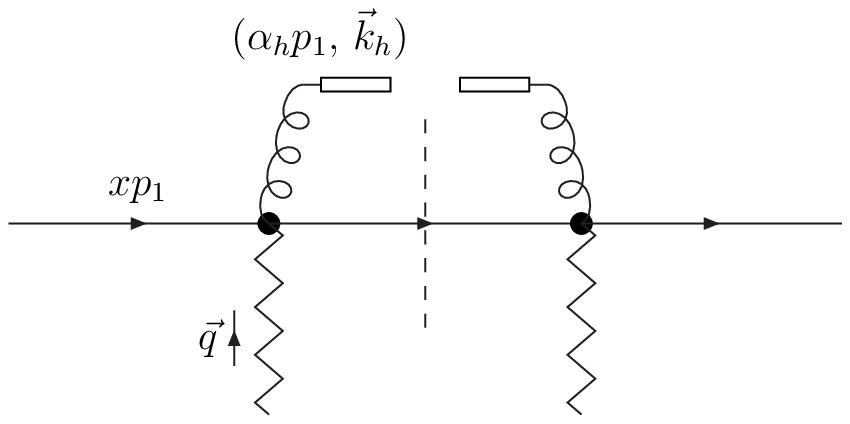}
\hspace{2cm}
\includegraphics[scale=0.6]{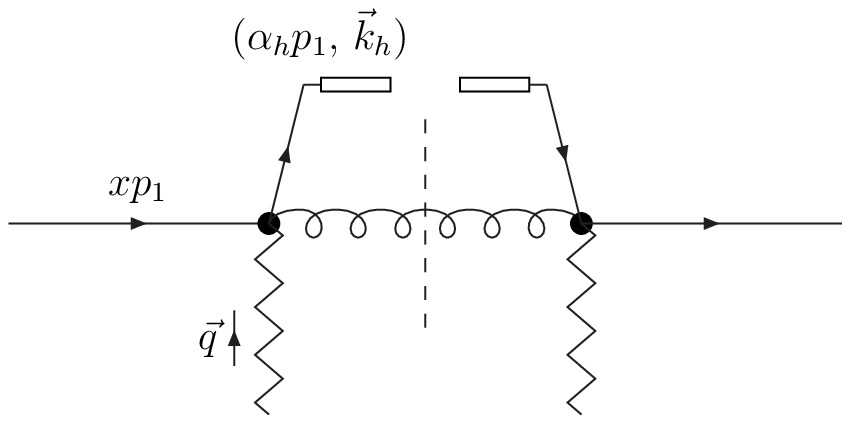}
\vspace{-0.2cm}
\caption[]{Diagrammatic representation of the NLA vertex for the case of 
incoming quark: real corrections from quark-gluon intermediate state, 
cases of gluon fragmentation (left) and quark fragmentation (right).}
\label{fig:vertex_quark_NLA_frag}
\end{figure}

For gluon fragmentation (see Fig.~\ref{fig:vertex_quark_NLA_frag} left), the 
``parent'' parton variables are $\vec k=\vec k_1$, $\zeta=\beta_1$ and
the contribution reads
\begin{eqnarray*}
&I_{q,g}^R=\frac{\alpha_s}{2\pi}\frac{1}{\epsilon}
\frac{\Gamma[1-\epsilon]}{(4\pi)^\epsilon}
\int\limits^1_{\alpha_h} \frac{dx}{x} \int\limits^1_{\frac{\alpha_h}{x}}
\frac{d\zeta}{\zeta} \sum_{a=q,\bar q}f_a(x)
D^h_g\left(\frac{\alpha_h}{x \zeta}\right)
\left(\vec k^{\,\,2} \right)^{\gamma+\epsilon-{n \over 2}}
\left(\vec k \cdot \vec l \,\, \right)^n & \\
& \times P_{gq}(\zeta)\left[\frac{C_A}{C_F}
+\zeta^{-2\gamma}\right]  + {\rm finite \ terms} &
\end{eqnarray*}

For quark fragmentation (see Fig.~\ref{fig:vertex_quark_NLA_frag} right), the
``parent'' parton variables are $\vec k=\vec k_2$, $\zeta=\beta_2$.
The contribution proportional to $C_F$ reads
\begin{eqnarray*}
&\left(I_{q,q}^R\right)^{C_F}=
\frac{\alpha_s}{2\pi}\frac{1}{\epsilon}
\frac{\Gamma[1-\epsilon]}{(4\pi)^\epsilon}\frac{\Gamma^2(1+\epsilon)}{\Gamma(1+2\epsilon)}
\int\limits^1_{\alpha_h} \frac{dx}{x} \int\limits^1_{\frac{\alpha_h}{x}}
\frac{d\zeta}{\zeta} \sum_{a=q,\bar q}f_a(x)
D^h_a\left(\frac{\alpha_h}{x \zeta}\right)
\left(\vec k^{\,\,2} \right)^{\gamma+\epsilon-{n \over 2}}
\left(\vec k \cdot \vec l \,\, \right)^n & \\
& \times \left\{
C_F \left(\frac{2}{\epsilon}-3\right)\delta(1-\zeta)
+P_{qq}(\zeta)\left(1+\zeta^{-2\gamma}\right)+ {\rm finite \ terms} \right\}\;,
&
\end{eqnarray*}
while the contribution proportional to $C_A$ reads
\begin{eqnarray*}
& \left(I_{q,q}^R\right)^{C_A}=\frac{\alpha_s}{2\pi}\frac{1}
{\epsilon}\frac{\Gamma(1-\epsilon)}{(4\pi)^\epsilon}
\int\limits^1_{\alpha_h} \frac{dx}{x}\int\limits^1_{\frac{\alpha_h}{x}}
\frac{d\zeta}{\zeta } \sum_{a=q,\bar q}
f_a(x) D_a^h\left(\frac{\alpha_h}{x\zeta }\right)
\left(\vec k^{\,\,2} \right)^{\gamma+\epsilon
-{n \over 2}}\left(\vec k \cdot \vec l \,\, \right)^n & \\
& \times  C_A \, \delta(1-\zeta)\ln\frac{s_0}{\vec k^2}
+ {\rm finite \ terms}\;.&
\end{eqnarray*}

\subsection{Gluon-initiated subprocess}

The virtual corrections to the one-particle intermediate state are
\[
I_g^V=-\frac{\alpha_s}{2\pi}\frac{\Gamma[1-\epsilon]}
{(4\pi)^\epsilon}\frac{1}{\epsilon}\frac{\Gamma^2(1+\epsilon)}
{\Gamma(1+2\epsilon)} \int\limits^1_{\alpha_h}\frac{dx}{x}
f_g(x) D_g^h\left(\frac{\alpha_h}{x}\right)\left(\vec k^{\,\,2} 
\right)^{\gamma+\epsilon-{n \over 2}}
\left(\vec k \cdot \vec l \,\, \right)^n \, \frac{C_A}{C_F}
\]
\[
\times \left\{
C_A\left(\ln\frac{s_0}{\vec k^2}+\frac{2}{\epsilon}-\frac{11}{6}\right)
+\frac{n_f}{3} \right\} + {\rm finite \ terms}\;.
\]

For the ``real'' corrections from quark-antiquark intermediate state, the 
starting point is the corresponding contribution to the ``inclusive'' gluon 
IF ($T_R=1/2$),
\[
\Phi^{\{Q\bar Q\}}\!=\!\Phi_g g^2\vec q^{\,\, 2}\frac{d^{2+2\epsilon} \vec k_1}
{(2\pi)^{3+2\epsilon}}d\beta_1 T_R\left(1-\frac{2\beta_1\beta_2}{1+\epsilon}
\right)\!\left\{\frac{C_F}{C_A}\frac{1}{\vec k_1^{\, 2} \vec k_2^{\, 2}}
+\beta_1\beta_2\frac{\vec k_1\cdot\vec k_2}{\vec k_1^{\, 2} \vec k_2^{\, 2}
(\vec k_2\beta_1-\vec k_1 \beta_2)^2} \right\},
\]
where $\beta_{1,2}$ and $\vec k_{1,2}$ are the relative longitudinal and 
transverse momenta of the quark (antiquark) and $\beta_1+\beta_2=1$, 
$\vec k_1+\vec k_2=\vec q$.

\begin{figure}[tb]
\centering
\includegraphics[scale=0.6]{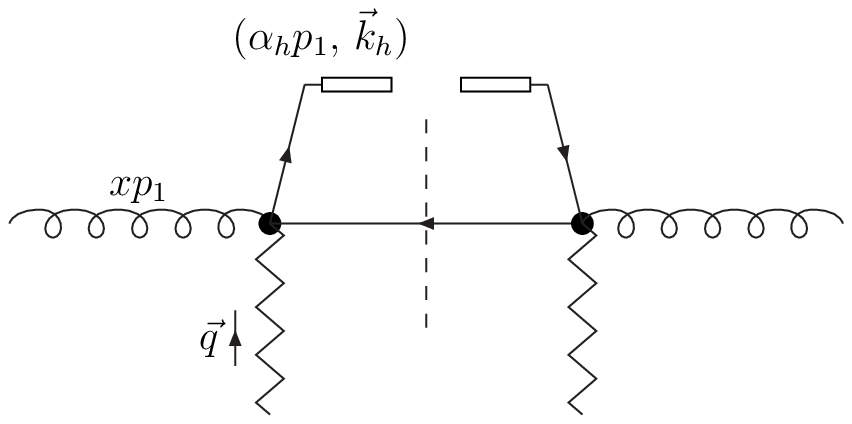}
\hspace{2cm}
\includegraphics[scale=0.6]{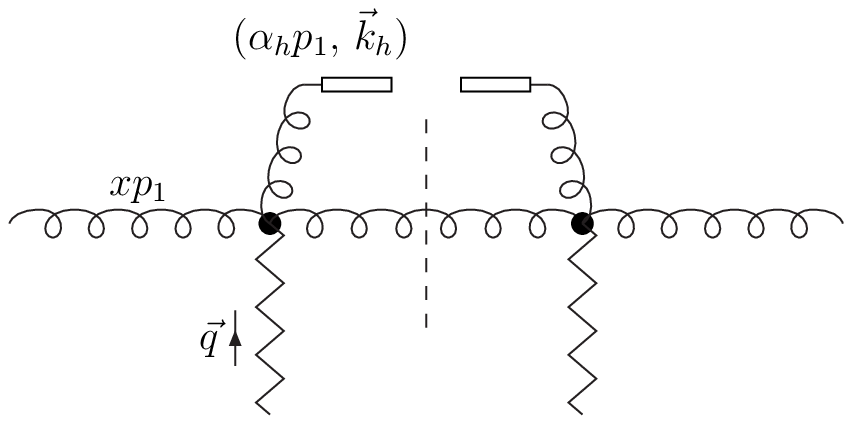}
\vspace{-0.2cm}
\caption[]{Diagrammatic representation of the NLA vertex for the case of 
incoming gluon: real corrections 
from quark-antiquark intermediate state, case of quark fragmentation (left)
and from two-gluon intermediate state, case of gluon fragmentation (right).}
\label{fig:vertex_gluon_NLA_frag}
\end{figure}

For quark (or antiquark) fragmentation (see 
Fig.~\ref{fig:vertex_gluon_NLA_frag} left) the ``parent'' parton variables are 
$\vec k=\vec k_1$, $\zeta=\beta_1$ and the contribution reads
\begin{eqnarray*}
&I^R_{g,q}=\frac{\alpha_s}{2\pi}\frac{1}{\epsilon}
\frac{\Gamma[1-\epsilon]}{(4\pi)^\epsilon}
\int\limits^1_{\alpha_h} \frac{dx}{x} \int\limits^1_{\frac{\alpha_h}{x}}
\frac{d\zeta}{\zeta} f_g(x) \sum_{a=q,\bar q}
D^h_a\left(\frac{\alpha_h}{x \zeta}\right)\left(\vec k^{\,\,2} 
\right)^{\gamma+\epsilon-{n \over 2}}
\left(\vec k \cdot \vec l \,\, \right)^n\, \frac{C_A}{C_F} & \\
& \times 
P_{qg}(\zeta)\left[\frac{C_F}{C_A}+\zeta^{-2\gamma}\right]
+ {\rm finite \ terms} \;.&
\end{eqnarray*}

For ``real'' corrections from the two-gluon intermediate state, the starting 
point is the corresponding contribution to the ``inclusive'' gluon IF,
\[
\Phi^{\{GG\}}=\Phi_g g^2\vec q^{\:2}\frac{d^{2+2\epsilon} \vec k_1}
{(2\pi)^{3+2\epsilon}}d\beta_1 \frac{C_A}{2}\left[\frac{1}{\beta_1}
+\frac{1}{\beta_2}-2+\beta_1\beta_2\right]
\]
\[
\times
\left\{\frac{1}{\vec k_1^{\, 2} \vec k_2^{\, 2}}+\frac{\beta_1^2}
{\vec k_1^{\, 2} (\vec k_2\beta_1-\vec k_1 \beta_2)^2}
+\frac{\beta_2^2}{\vec k_2^{\, 2} (\vec k_2\beta_1-\vec k_1 \beta_2)^2}
\right\}\;,
\]
where $\beta_{1,2}$ and $\vec k_{1,2}$ are the relative longitudinal and 
transverse momenta of the two gluons and $\beta_1+\beta_2=1$, 
$\vec k_1+\vec k_2=\vec q$. We can have only gluon fragmentation to be counted 
with a factor of 2 (see Fig.~\ref{fig:vertex_gluon_NLA_frag} right). The
result is
\begin{eqnarray*}
&I^R_{g,g}=\frac{\alpha_s}{2\pi}\frac{1}{\epsilon}
\frac{\Gamma[1-\epsilon]}{(4\pi)^\epsilon}
\frac{\Gamma^2(1+\epsilon)}{\Gamma(1+2\epsilon)}
\int\limits^1_{\alpha_h} \frac{dx}{x}  \int\limits^1_{\frac{\alpha_h}{x}}
\frac{d\zeta}{\zeta} f_g(x) D^h_g\left(\frac{\alpha_h}
{x \zeta}\right)\left(\vec k^{\,\,2} \right)^{\gamma+\epsilon-{n \over 2}}
\left(\vec k \cdot \vec l \,\, \right)^n\, \frac{C_A}{C_F}& \\
&\times \left\{
P_{gg}(\zeta)\left(1+\zeta^{-2\gamma}\right)+
\delta(1-\zeta)\left[C_A\left(\ln\frac{s_0}{\vec k^2}
+\frac{2}{\epsilon}-\frac{11}{3}\right)+\frac{2n_f}{3}\right]\right\}
+ {\rm finite \ terms} \;.&
\end{eqnarray*}

\section{Final result and discussion}

One can verify that all UV and IF divergences cancel, leading to
\[
\vec k^{\, 2}_h \,\,\frac{d\Phi^h(\nu,n)}
{d\alpha_h d^2\vec k_h}=2\,\alpha_s(\mu_R)\sqrt{\frac{C_F}{C_A}}
\left(\vec k_h^{\,\,2}
\right)^{\gamma-{n \over 2}}\!\left(\vec k_h \cdot \vec l \, \right)^n
\!\left\{
\int\limits^1_{\alpha_h}\frac{dx}{x} \left(\frac{x}{\alpha_h}\right)^{2\gamma} 
\!\left[ \frac{C_A}{C_F}f_g(x)
D_g^h\left(\frac{\alpha_h}{x}\right) \right.\right.
\]
\[
\left. +\sum_{a=q,\bar q}f_a(x)
D_a^h\left(\frac{\alpha_h}{x}\right)\right]
+ \frac{\alpha_s\left(\mu_R\right)}{2\pi}\int\limits^1_{\alpha_h} \frac{dx}{x} 
\int\limits^1_{\frac{\alpha_h}{x}}
\frac{d\zeta}{\zeta}\left(\frac{x\, \zeta}{\alpha_h}\right)^{2\gamma}
\]
\[
\times
\left[\frac{C_A}{C_F}f_g(x)
D_g^h\left(\frac{\alpha_h}{x\zeta}\right)
C_{gg}\left(x,\zeta\right)+\sum_{a=q,\bar q}f_a(x)
D_a^h\left(\frac{\alpha_h}{x\zeta}\right)
C_{qq}\left(x,\zeta\right) \right.
\]
\[
\left.\left.
+\sum_{a=q,\bar q}f_a(x)
D_{g}^h\left(\frac{\alpha_h}{x\zeta}\right)
C_{qg}\left(x,\zeta\right)
+\frac{C_A}{C_F} f_{g}(x)\sum_{a=q,\bar q}
D_a^h\left(\frac{\alpha_h}{x\zeta}\right)
C_{gq}\left(x,\zeta\right) \right]
\right\}\;.
\]
The explicit form of the coefficient functions will be given elsewhere
~\cite{IP}.

To summarize, we have discussed the NLA calculation of the IF the 
forward production of an identified hadron from an incoming quark or gluon, 
emitted by a proton. This is a necessary ingredient for the hard inclusive 
production of a pair of rapidity-separated identified hadrons at LHC.
We have given our result in the $(\nu,n)$-representation, which is the most 
convenient for the numerical determination of the cross section~\cite{mesons}.
We have shown that soft and virtual infrared 
divergences cancel each other, whereas the IR collinear ones are 
compensated by the PDFs' and FFs' renormalization counterterms, the remaining 
UV divergences being taken care of by the QCD coupling renormalization.

\vspace{0.2cm} 

{\bf Acknowledgments.} D.I. thanks the Dipartimento di Fisica 
dell'Universit\'a della Calabria and the Istituto Nazionale di Fisica 
Nucleare (INFN), Gruppo collegato di Cosenza, for the warm hospitality
and the financial support. This work was also supported in part by the
grants RFBR-09-02-00263 and RFBR-11-02-00242.

\end{document}